\documentclass[sigconf, anonymous=false]{acmart}

\usepackage{cleveref}
\usepackage[color=lightgray]{todonotes}
\usepackage[ruled, vlined, linesnumbered]{algorithm2e}
\usepackage{amsfonts}
\usepackage{subcaption}
\usepackage{multirow}
\usepackage{bm}
\usepackage{soul}
\usepackage{bbold}
\usepackage{mathtools}
\usepackage[utf8]{inputenc}
\usepackage{graphicx}
\usepackage{booktabs}
\usepackage{listings}
\usepackage{xcolor}
\usepackage{tabularx}
\usepackage{makecell}
\usepackage{pifont}
\usepackage{siunitx}
\newcommand{\cmark}{\ding{51}}%
\newcommand{\xmark}{\ding{55}}%
\crefname{lstlisting}{algorithm}{algorithms}
\Crefname{lstlisting}{Algorithm}{Algorithms}

\newcommand{\xhdr}[1]{\vspace{1.5ex}\noindent\textbf{#1}}

\definecolor{codegreen}{rgb}{0,0.6,0}
\definecolor{codegray}{rgb}{0.5,0.5,0.5}
\definecolor{codepurple}{rgb}{0.58,0,0.82}
\definecolor{backcolour}{rgb}{0.95,0.95,0.92}

\lstdefinestyle{mystyle}{
    commentstyle=\color{codegreen},
    keywordstyle=\color{magenta},
    numberstyle=\tiny\color{codegray},
    stringstyle=\color{codepurple},
    basicstyle=\ttfamily\footnotesize,
    breakatwhitespace=false,         
    breaklines=true,                 
    captionpos=b,                    
    keepspaces=true,                 
    showspaces=false,                
    showstringspaces=false,
    showtabs=false,                  
    tabsize=2,
}
\newcommand{\ours}{{LRURec}\xspace}
\AtBeginDocument{%
  \providecommand\BibTeX{{%
    \normalfont B\kern-0.5em{\scshape i\kern-0.25em b}\kern-0.8em\TeX}}}

\begin{document}


\title{Linear Recurrent Units for Sequential Recommendation}

\author{Zhenrui Yue}
\authornote{Both authors contributed equally to this research.}
\email{zhenrui3@illinois.edu}
\affiliation{%
  \institution{University of Illinois Urbana-Champaign}
  \city{Champaign}
  \country{USA}
}

\author{Yueqi Wang}
\authornotemark[1]
\email{yueqi@berkeley.edu}
\affiliation{%
  \institution{University of California, Berkeley}
  \city{Berkeley}
  \country{USA}
}

\author{Zhankui He}
\authornote{Correspondence to Zhankui He (\url{zhh004@ucsd.edu}).}
\email{zhh004@ucsd.edu}
\affiliation{%
  \institution{University of California, San Diego}
  \city{San Diego}
  \country{USA}
}

\author{Huimin Zeng}
\email{huiminz3@illinois.edu}
\affiliation{%
  \institution{University of Illinois Urbana-Champaign}
  \city{Champaign}
  \country{USA}
}

\author{Julian McAuley}
\email{jmcauley@ucsd.edu}
\affiliation{%
  \institution{University of California, San Diego}
  \city{San Diego}
  \country{USA}
}

\author{Dong Wang}
\email{dwang24@illinois.edu}
\affiliation{%
  \institution{University of Illinois Urbana-Champaign}
  \city{Champaign}
  \country{USA}
}

\renewcommand{\shortauthors}{Yue and Wang, et al.}

\begin{abstract}
State-of-the-art sequential recommendation relies heavily on self-attention-based recommender models. Yet such models are computationally expensive and often too slow for real-time recommendation. Furthermore, the self-attention operation is performed at a sequence-level, thereby making low-cost incremental inference challenging. Inspired by recent advances in efficient language modeling, we propose \ul{l}inear \ul{r}ecurrent \ul{u}nits for sequential \ul{rec}ommenda-tion (\ours). Similar to recurrent neural networks, \ours offers rapid inference and can achieve incremental inference on sequential inputs. By decomposing the linear recurrence operation and designing recursive parallelization in our framework, \ours provides the additional benefits of reduced model size and parallelizable training. Moreover, we optimize the architecture of \ours by implementing a series of modifications to address the lack of non-linearity and improve training dynamics. To validate the effectiveness of our proposed \ours, we conduct extensive experiments on multiple real-world datasets and compare its performance against state-of-the-art sequential recommenders. Experimental results demonstrate the effectiveness of \ours, which consistently outperforms baselines by a significant margin. Results also highlight the efficiency of \ours with our parallelized training paradigm and fast inference on long sequences, showing its potential to further enhance user experience in sequential recommendation.
\end{abstract}






\maketitle

\section{Introduction}
\label{sec:intro}

\begin{figure}[t]
    \centering
    \includegraphics[trim=1.8cm 0 3cm 0, clip, width=1.0\linewidth]{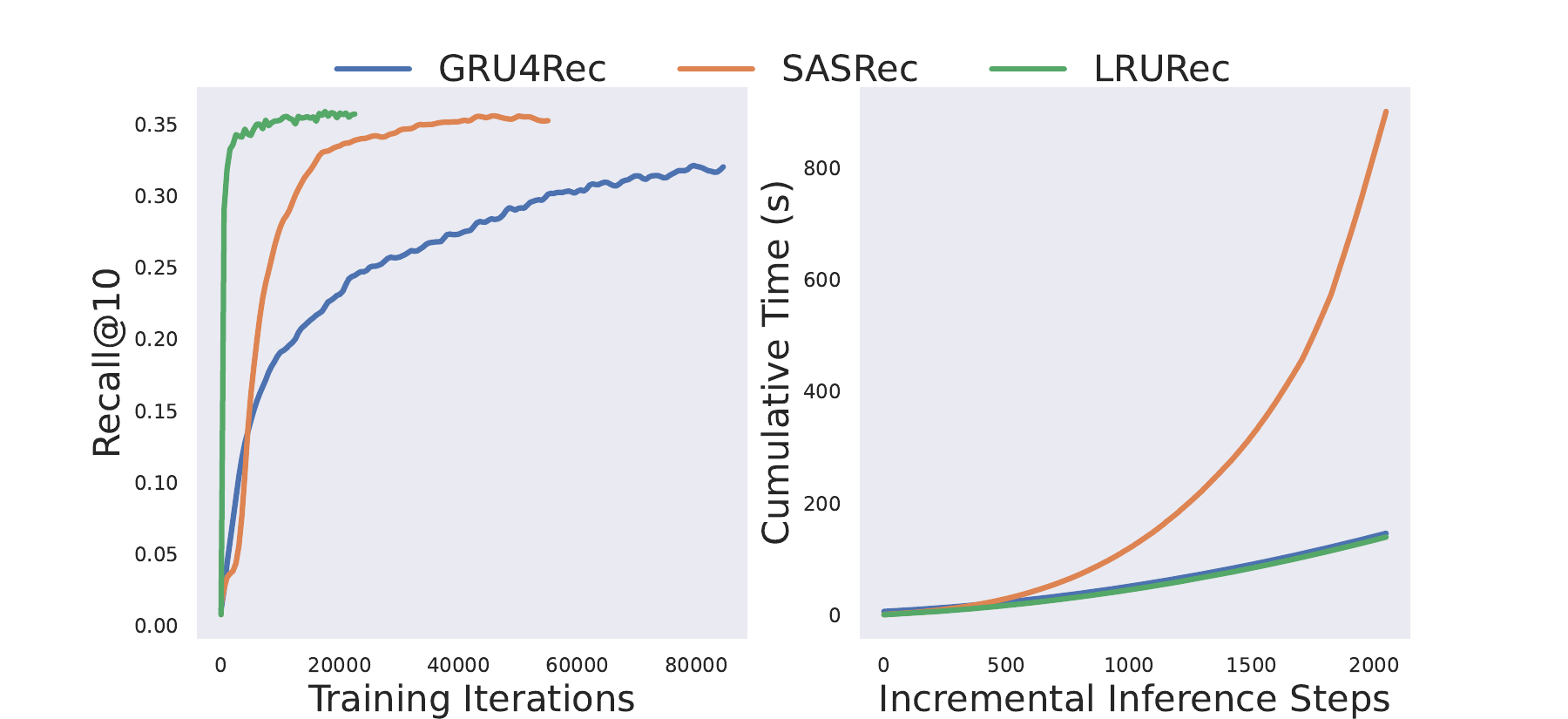}
    \caption{Training and inference efficiency of GRU4Rec, SASRec and \ours on ML-1M. The proposed \ours converges significantly faster than SASRec and GRU4Rec, while outperforming both models on Recall@10 scores and achieving $\bm{\mathcal{O}(1)}$ complexity with incremental inference.}
    \label{fig:training-inference}
    \vspace{-10pt}
\end{figure}

Sequential recommender systems play a crucial role in personalization platforms, such as Netflix~\cite{steck2021deep} and Amazon~\cite{li2022coarse}, by capturing sequential patterns from user action history (e.g., user clicks) to accurately predict the next action. Over time, these recommenders have undergone significant advances from traditional approaches like Markov chain~\cite{rendle2010factorizing,he2016fusing} to deep neural networks like convolutional neural networks (CNNs)~\cite{tang2018personalized,yan2019cosrec} and recurrent neural networks (RNNs)~\cite{hidasi2015session, hidasi2018recurrent,li2017neural}. Recently, self-attentive recommenders (SARs), inspired by transformers and their variants~\cite{vaswani2017attention,devlin-etal-2019-bert}, have further improved both training efficiency and recommendation accuracy, and thus represent the current state-of-the-art~\cite{kang2018self,sun2019bert4rec,he2021locker}.

Despite their superior performance and training efficiency, SARs face criticism from the recommendation community due to their efficiency issues~\cite{pancha2022pinnerformer}. Specifically, SARs~\cite{kang2018self,sun2019bert4rec,he2021locker} heavily rely on computing item-to-item weights (referred to as ``self-attention'') at a sequence-level to encode user representations over time. That is, each time a new user action occurs, the system appends the action to the user history, and then recomputes all item-to-item weights for incremental update on user representation. This process incurs significant time and space costs. In contrast, RNNs~\cite{hidasi2015session, hidasi2018recurrent,li2017neural} differ from the following perspectives: (1)~in terms of inference, RNNs perform efficiently by storing a fixed-length hidden-state vector as the user representation. RNNs can retrieve and update the vector step-wise whenever a new user action occurs. (2)~RNNs face challenges in training efficiency. The recurrent non-linear units in RNNs prevent parallelizable training, leading to the need for cumbersome training sequence augmentation and slow convergence speed. (3)~Moreover, empirical studies have shown that existing RNNs fail to match the recommendation performance of SARs~\cite{kang2018self,sun2019bert4rec,he2021locker}.

In this work, we propose a novel sequential recommender model, \ul{l}inear \ul{r}eccurent \ul{u}nits for sequential \ul{rec}ommendation (\ours), which not only outperforms the recommendation accuracy and training efficiency of SARs with parallelizable training (see~\Cref{fig:training-inference}), but also matches the inference efficiency of RNNs by only maintaining a fixed-length user vector for incremental updates. Our motivation stems from two key observations: (1)~the effectiveness of linear models in handling long sequences, as demonstrated in natural language processing (NLP) tasks and models~\cite{peng2023rwkv,orvieto2023resurrecting,sun2023retentive}; and (2)~our findings that linear RNN recommenders can achieve recommendation performance comparable to vanilla non-linear RNN recommenders. To begin, we introduce LRU~\cite{gu2021efficiently,orvieto2023resurrecting}, a linear version of the traditional recurrent units, which supports parallelizable training with the proposed recursive parallelization. This enables more efficient training compared to its non-linear counterparts. Furthermore, we incorporate a series of sequential modeling techniques from self-attentive architectures, such as layer normalization~\cite{ba2016layer}, residual connection~\cite{he2016deep}, and position-wise feed-forward networks~\cite{vaswani2017attention}. They further improve the recommendation accuracy to outperform existing SARs while retaining RNN-like inference efficiency due to the replacement of self-attention computation with efficient RNN-like recurrence operation.

We summarize the contributions of our work as follows\footnote{Our implementation is available at https://github.com/yueqirex/LRURec.}:
\begin{itemize}
    \item We introduce a linear recurrent unit-based architecture \ours into sequential recommendation, which addresses the dilemma of training efficiency, inference efficiency and recommendation performance.
    \item We propose a series of improvements in \ours to address the lack of non-linearity and improve training dynamics. We further propose recursive parallelization that significantly accelerates both training and inference.
    \item We empirically demonstrate the effectiveness and efficiency of \ours in comparison to state-of-the-art methods 
    on multiple real-world datasets, where \ours consistently outperforms baseline methods by a large margin.
    \item Our results challenge the necessity of the core self-attention module in existing SARs while highlighting the importance of other techniques in SARs like layer normalization, which provide deeper understanding and new opportunities to the architecture design for sequential recommenders.
\end{itemize}

\begin{table}[t]
\small
\begin{tabular}{lcccc}
\toprule
\textbf{Group}                   & \textbf{Model} & \makecell{\textbf{Training} \\\textbf{Efficiency}} & \makecell{\textbf{Inference} \\ \textbf{Efficiency}} & \makecell{\textbf{Rec.} \\ \textbf{Performance}} \\ \midrule
\multirow{2}{*}{\textbf{RNNs}}   & GRU4Rec~\cite{hidasi2015session}             &    \xmark            &    \cmark                                          &    \cmark                                     \\
                                 & NARM~\cite{li2017neural}                     &    \xmark            &    \cmark                                          &    \cmark                                     \\ \midrule
\multirow{2}{*}{\textbf{SARs}}   & SASRec~\cite{kang2018self}                   &    \cmark            &    \xmark                                          &    \cmark\cmark                               \\
                                 & BERT4Rec~\cite{sun2019bert4rec}              &    \cmark            &    \xmark                                          &    \cmark\cmark                               \\ \midrule
\multirow{2}{*}{\textbf{Others}} & FPMC~\cite{rendle2010factorizing}            &    \xmark            &    \cmark                                          &    \cmark                                     \\
                                 & FMLP-Rec~\cite{zhou2022filter}               &    \cmark            &    \xmark                                          &    \cmark\cmark                               \\ \midrule
\textbf{Ours}                    & \ours                                        &    \cmark            &    \cmark                                          &    \cmark\cmark                               \\ \bottomrule
\end{tabular}
\caption{Comparison among the proposed \ours and representative sequential recommenders. \ours is both training-efficient (like SARs~\cite{kang2018self, sun2019bert4rec}) and inference-efficient (like RNNs~\cite{hidasi2015session, li2017neural} or FPMC~\cite{rendle2010factorizing}). Additionally, \ours matches SARs and FMLP-Rec~\cite{zhou2022filter} on recommendation performance.} 
\label{tab:intro_comparison}
\end{table}
\section{Related Work}

\subsection{Sequential Recommendation}

Sequential Recommendation has a primary objective of predicting the user's next item of interest by modeling their past actions in chronological order~\cite{rendle2010factorizing,hidasi2015session,tang2018personalized,kang2018self}. The central focus of sequential recommendation lies in the creation of a model architecture that is both effective and efficient. First, CNN- and RNN-based models were proposed to leverage the expressiveness of deep neural networks to model user sequences~\cite{hidasi2015session,tang2018personalized,li2017neural}, which outperform linear sequential models such as FMC and FPMC~\cite{rendle2010factorizing}. Nevertheless, RNN-based models faced limitations in training efficiency due to the lack of parallel training. Subsequently, transformer-based models~\cite{kang2018self,sun2019bert4rec,he2021locker,he2022query} emerged as a solution, offering accelerated training and improved recommendation accuracy through the parallelizable self-attention operation~\cite{vaswani2017attention}. With efficient training and accurate item-to-item relevance via self-attention, transformer-based models stand as the \emph{state-of-the-art} methods for sequential recommendation. Additionally, MLP-based sequential recommenders~\cite{li2022mlp4rec,li2023automlp} like FMLP-Rec~\cite{zhou2022filter} exhibit comparable recommendation performance relying solely on the MLP architecture. However, the inference efficiency of transformer- and MLP-based models lag behind RNN-based models~\cite{pancha2022pinnerformer}. Specifically, whenever a new user action appears, RNN-based models only need to update the latest hidden state, whereas both transformer- and MLP-based models must recompute sequence-level relevance for inference. 

In this work, we aim to propose a new model architecture for sequential recommendation based on linear recurrent units~\cite{orvieto2023resurrecting}, which enjoys both the training efficiency of transformer-based models and inference efficiency of RNN-based models.

\subsection{Efficient Language Modeling}

The difficulty of building an NLP model to achieve (1)~training efficiency, (2)~inference efficiency and (3)~model performance like~\Cref{tab:intro_comparison} is known as the ``impossible triangle''~\cite{sun2023retentive}. To achieve these goals, various approaches have been proposed from different aspects. Attention free transformer (AFT) or RWKV~\cite{zhai2021attention, peng2023rwkv} simplify token-token attention weights to element-wise operations and moves softmax operations to key vectors, which shows comparable performance to a vanilla transformer model~\cite{vaswani2017attention} after scaling up~\cite{peng2023rwkv}. S4~\cite{gu2021efficiently} is based on fundamental state space models (SSM) with continuous-time memorization~\cite{gu2021combining}, which demonstrates efficiency and effectiveness in long-sequence modeling (e.g.~long-range arena benchmark~\cite{tay2020long}). Linear recurrent units~\cite{orvieto2023resurrecting} match the long-sequence performance of SSMs with an improved initialization strategy and systematic ablations. Most recently, RetNet~\cite{sun2023retentive} proposes attention-like retention that have equivalent recurrent and parallel formulation, RetNet can thus be trained in parallel while achieving low inference costs like RNN models.

In our work, we study the efficient modeling problem in the context of sequential recommendation. The differences are (1)~all mentioned methods are primarily studied and evaluated in NLP, where existing solutions (e.g., RWKV) are not specifically designed for recommendation and can lead to overfitting and latency issues; (2)~the majority of such methods are optimized for long-range modeling tasks, which may cause performance issues upon recommendation tasks that prioritize short sequences of user actions. Different from existing approaches, we propose a novel sequential recommender based on linear recurrence. With the carefully designed LRU block and recursive parallelization, our model performs well regardless of sequence length and achieves both training and inference efficiency for sequential recommendation.

\section{Methodology}

As shown in previous works~\cite{gu2021efficiently,orvieto2023resurrecting,sun2023retentive}, efficient NLP models are capable of capturing long-term dependencies in a wide range of sequence-to-sequence tasks. Additionally, such models demonstrate significantly improved efficiency thanks to parallelizable training and incremental inference. Motivated by such observations, the proposed linear recurrent units for sequential recommendation (\ours) incorporates the advantages of both RNN-based and transformer-based models, while requiring significantly reduced computing power with efficient incremental inference and rapid convergence via parallelized training. We introduce the key components and the overall model of \ours in the following.

\subsection{Setup}

\xhdr{Data}: Let input $x \in \mathcal{X}$ represent the user action history $[x_1, x_2, ..., x_L]$ of length $L$ in chronological order, where the elements are represented in the item space $\mathcal{I}$ (i.e., $x_i \in \mathcal{I}$). The ground truth $y$ is the next user action $x_{L+1} \in \mathcal{I}$ (i.e., $y = x_{L+1}$).

\xhdr{Model}: The recommender model is denoted with function $f$, with $H$ being the hidden dimension in $f$. Upon input $x$, $f$ generates prediction scores over $\mathcal{I}$. Ideally, $f$ predicts item $y$ with the highest score for data pair $(x, y)$, namely $y = \arg\max f(x)$. 

\xhdr{Learning}:
The optimization of the recommender model corresponds to the likelihood maximization of ground truth item $y$ upon input $x$. In other words, the learning objective is to minimize the negative log-likelihood loss $\mathcal{L}$ over distribution $\mathcal{X}$:
\begin{equation}
    \min \mathbb{E}_{(x, y) \sim \mathcal{X}} \mathcal{L}(f(x), y).
\end{equation}

\subsection{Linear Recurrent Unit}
\label{sec:lru}

\xhdr{Decomposing Linear Recurrence.}
We first introduce a simplified form of linear recurrence~\cite{gu2021efficiently}. For input $x_k$ at time step $k$, we represent the hidden representation $h_k$ and output $y_k$ with learnable matrices $A \in \mathbb{R}^{H \times H}$, $B \in \mathbb{R}^{H \times H_{\mathrm{in}}}$, $C \in \mathbb{R}^{H_{\mathrm{out}} \times H}$ and $D \in \mathbb{R}^{H_{\mathrm{out}} \times H_{\mathrm{in}}}$:
\begin{equation}
    h_k = A h_{k-1} + B x_k, \quad y_k = C h_k + D x_k,
    \label{eq:linear-recurrence}
\end{equation}
where the input and output dimensions are denoted with $H_{\mathrm{in}}$ and $H_{\mathrm{out}}$ (i.e., embedding size), and the hidden dimension size with $H$. Different from RNN models (i.e., $h_k = \sigma(A h_{k-1} + B x_k)$), we discard the non-linearity $\sigma$ to enable the serialization of $h_k$:
\begin{equation}
  \begin{aligned}
    h_k &= A h_{k-1} + B x_k = A^2 h_{k-2} + A B x_{k-1} + B x_k \\
        &= A^3 h_{k-3} + A^2 B x_{k-2} + A B x_{k-1} + B x_k = \ldots \\
        &= \sum_{i=1}^{k} A^{k-i} B x_{i} \quad \mathrm{with} \quad h_1 = B x_1.
  \label{eq:unrolled-linear-recurrence}
  \end{aligned}
\end{equation}
By unrolling the $h_k = A h_{k-1} + B x_k$ along the time steps, $h_k$ can be written in closed-form w.r.t.~the matrices $A$, $B$ and the input elements $x_1, x_2, \ldots, x_k$. However, the repeated computation of matrix multiplication is inefficient and may lead to numerical issues as $k$ increases (e.g., overflow). To this end, we leverage matrix diagonalization (i.e., eigendecomposition) and introduce eigenvalues and eigenvectors to (1)~reduce matrix-level computation and (2)~control the numerical stability of $h_k$. In particular, we perform decomposition on $A$ with $A = P \Lambda P^{-1}$, in which $\Lambda$ is diagonal and consists of the eignevalues $\lambda_1, \lambda_2, \ldots, \lambda_{H}$, $P$ is an invertible matrix of size $H \times H$. Consequently, the computation of $A^{n}$ reduces to $P \Lambda^n P^{-1}$, and thereby significantly improving  computation efficiency.

\xhdr{Representation in Complex Space.}
Despite the eigendecomposition of matrix $A = P \Lambda P^{-1}$, the eigenvalues and eigenvectors of $A$ do not necessarily lie in real space $\mathbb{R}$. Therefore, we extend $P$ and $\Lambda$ to the complex space with $P \in \mathbb{C}^{H \times H}$ and $\Lambda = \mathrm{diag}(\lambda_1, \lambda_2, \ldots, \lambda_{H}) \in \mathbb{C}^{H \times H}$~\cite{orvieto2023resurrecting}. Similarly, we extend $h$ and $B$, $C$ to the complex space $\mathbb{C}$. As such, the computation of $\sum_{i=1}^{k} A^{k-i} B x_i$ can now be written as $\sum_{i=1}^{k} P \Lambda^{k-i} P^{-1} B x_{i}$. We further write $\bar{h} = P^{-1} h$, $\bar{B} = P^{-1} B$ and $\bar{C} = C P^{-1}$, which simplifies the formulation of $\bar{h}_k$ and output $y_k$ to:
\begin{equation}
  \begin{aligned}
    \bar{h}_k &= \Lambda \bar{h}_{k-1} + \bar{B} x_k = \ldots = \sum_{i=1}^{k} \Lambda^{k-i} \bar{B} x_{i}, \\
    y_k &= \Re ( \bar{C} \bar{h}_k ) + D x_k,
  \label{eq:decomposed-linear-recurrence}
  \end{aligned}
\end{equation}
where $\Re ( \bar{C} \bar{h}_k )$ is the real part of complex vector $\bar{C} \bar{h}_k$. To improve multiplication efficiency in the complex space, we represent $\Lambda$ in polar form with absolute value $r$ and argument $\theta$ (i.e., $\Lambda = r e^{j\theta} = r(\mathrm{cos}(\theta) + j\mathrm{sin}(\theta))$), $j$ stands for the imaginary unit. Let $r = e^{-\nu}$, we have $\Lambda = r e^{j\theta} = e^{-\nu + j\theta}$ where $\nu, \theta \in \mathbb{R}^{H}$, thus the involved computation is reduced to $(-\nu + j\theta)$. Another advantage of the polar form is that parameters $\nu$ and $\theta$ can be now optimized in the real space $\mathbb{R}$. As a result, the scale of $\Lambda$ can be computed with $e^{-\nu}$, while the exponentiation of matrix $A$ in \Cref{eq:unrolled-linear-recurrence} is replaced with the efficient exponential of diagonal matrix $\Lambda$.

\xhdr{Parameterization.}
To avoid numerical instability, a simple condition is to let elements in $\Lambda$ satisfy $| \lambda_i | < 1, i = 1, 2, \ldots, H$, the condition is equivalent to $e^{-\nu_i} < 1$ for the polar form. Therefore, we use another exponential $\lambda_i = \exp (-\exp (\log(\nu_i))+ j\theta_i )$ with $\nu_i > 0$~\cite{gu2021efficiently, orvieto2023resurrecting}. More specifically, we use log parameters $\nu^{\log}$, $\theta^{\log} \in \mathbb{R}^{H}$ and introduce normalization factor $\gamma^{\log} \in \mathbb{R}^{H}$, with $\nu^{\log} = \log(\nu), \theta^{\log} = \log(\theta)$. $\gamma^{\log}$ normalizes input element-wise and is initialized with $\gamma^{\log}_i \leftarrow \log(\sqrt{1 - | \lambda_i |^2})$. The rest log parameters are initialized following the ring approach with radius between $[0.8, 0.99]$~\cite{orvieto2023resurrecting}, while the rest matrices are initialized via truncated normal initialization~\cite{sun2019bert4rec}. Using $h$, $B$ and $C$ to represent $\bar{h}$, $\bar{B}$ and $\bar{C}$ in the complex space $\mathbb{C}$, we summarize the above parameterization and formulate the final linear recurrence unit as follows:
\begin{equation}
  \begin{aligned}
    \Lambda_{\;} &= \mathrm{diag}(\exp({-\exp(\nu^{\log}) + j\exp(\theta^{\log})})), \\
    h_k &= \Lambda h_{k-1} + \exp(\gamma^{\log}) \odot B x_k, \\
    y_k &= \Re ( C h_k ) + x_k, \; \mathrm{with} \; h_1 = B x_1.
  \label{eq:finalized-linear-recurrence}
  \end{aligned}
\end{equation}
In our implementation, we double the size of $H$ in $\Lambda$ to improve the modeling of recurrence. We also adopt identity matrix $\mathbb{I}$ as $D$ in the output function as $H_{\mathrm{in}} = H_{\mathrm{out}}$. Discarding $D$ additionally reduces learnable parameters and constructs a residual connection between the input and output. In the following, we use \texttt{LRU} to denote the proposed linear recurrence operation in \Cref{eq:finalized-linear-recurrence}, the effectiveness of our design is demonstrated in \Cref{sec:ablation}.

\subsection{Recursive Parallelization}
\label{sec:parallelization}
Although matrix diagonalization and the exponential parameterization improves the computational efficiency of linear recurrence, the forward pass in \Cref{eq:finalized-linear-recurrence} is element-wise for each time step, resulting in linear time complexity w.r.t. sequence length. Inspired by the parallel scan algorithms~\cite{ladner1980parallel, blelloch1989scans, blelloch1990prefix, kexuefm-9554}, we develop recursive parallelization designed specifically for \ours. In particular, long input sequences are recursively split into subsequences to enable parallel processing, followed by the scaled addition of hidden features from each subsequence, and thereby improving the forward pass of \ours to logarithmic time complexity.

\begin{figure}[t]
    \centering
    \includegraphics[trim=5.8cm 2.85cm 5.7cm 2.2cm, clip, width=0.91\linewidth]{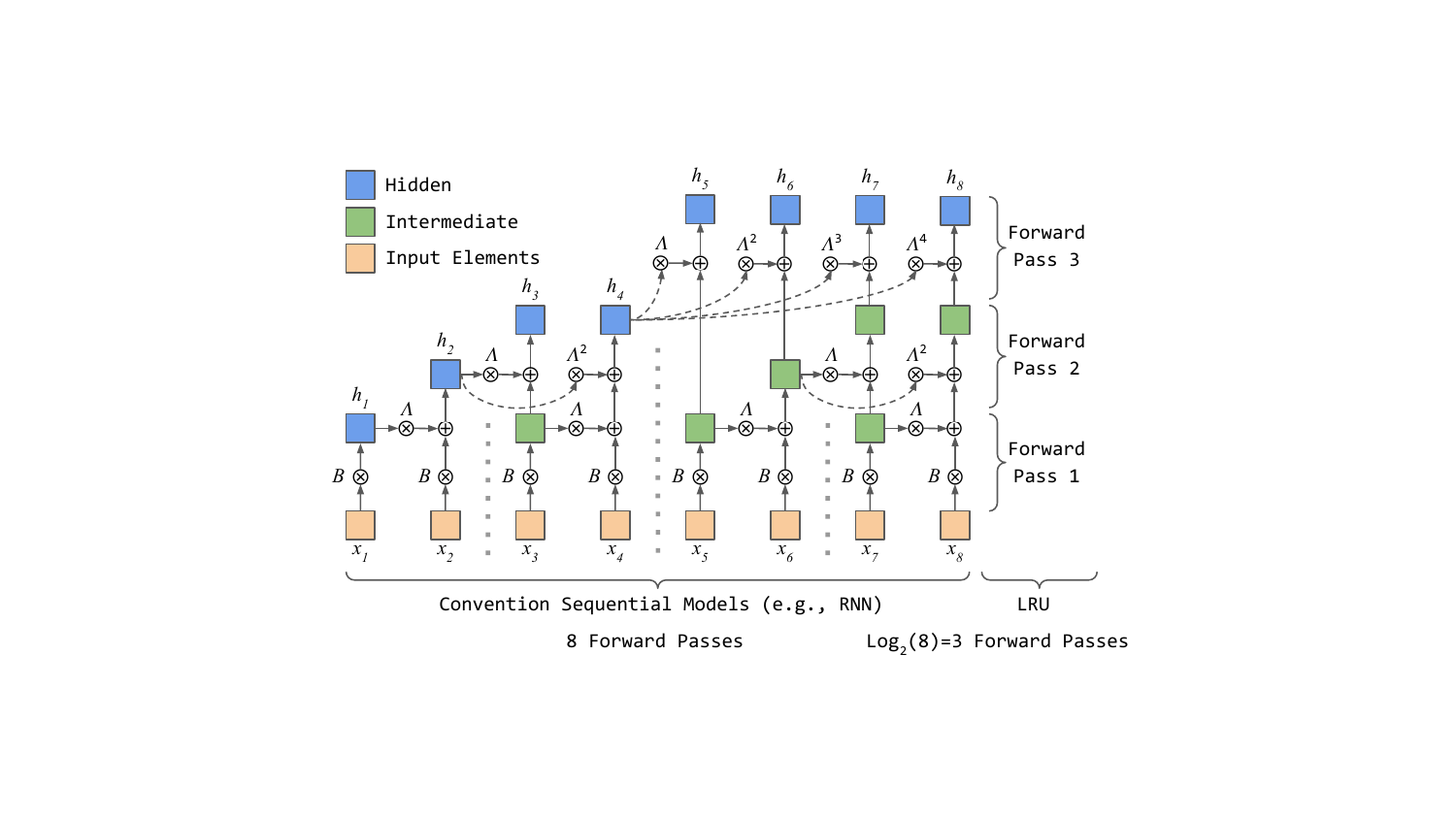}
    \caption{Recursive parallelization for \ours, we illustrate the recursive split and parallel forward pass.}
    \label{fig:parallel}
\end{figure}

For simplicity, we formulate the recursive parallelization based on $h_t = \sum_{i=1}^{k} \Lambda^{k-i} B x_{i}$ in \Cref{eq:decomposed-linear-recurrence}. Given input sequence $x$ of length $t$ and time step $k$ with $k < t$, the final output $h_t$ at time step $t$ can be formulated as:
\begin{equation}
  \begin{aligned}
    h_t &= \sum_{i=1}^{t} \Lambda^{t-i} B x_{i} = \sum_{i=1}^{k} \Lambda^{t-i} B x_{i} + \sum_{i=k+1}^{t} \Lambda^{t-i} B x_{i} \\
        &= \Lambda^{t-k} \sum_{i=1}^{k} \Lambda^{k-i} B x_{i} + \sum_{i=k+1}^{t} \Lambda^{t-i} B x_{i} \\
        &= \Lambda^{t-k} h_k + \sum_{i=k+1}^{t} \Lambda^{t-i} B x_{i}.
  \label{eq:recursive-linear-recurrence}
  \end{aligned}
\end{equation}
The results suggest that it is possible to split the sequence $x = [x_1, x_2, ..., x_t]$ into $[x_1, x_2, ..., x_k]$ and $[x_{k+1}, x_2, ..., x_t]$ for parallel processing on the subsequences. Then, $h_t$ can be obtained by summing the output from the subsequences, where the last-step hidden feature from the first subsequence (i.e., $h_k$) is additionally multiplied with $\Lambda^{t-k}$ to correct the time steps.

Based on such observations, we propose recursive parallelization to accelerate the forward feeding of sequential input. The recursive parallelization process is illustrated in \Cref{fig:parallel}. Specifically, we perform the the following steps in recursive parallelization:

\begin{lstlisting}[float=t, caption={\ours Recursive Parallelization, with $\mathtt{x}$, $\mathtt{B}$ and $\mathtt{La}$ representing input $x$, parameters $B$ and $\Lambda$.}, label={lst:parallel}, language=Python, belowskip=-1.0\baselineskip]
def recursive_parallelization(x, B, La):
    # 1. sequence padding
    L_log2 = int(math.ceil(math.log2(x.shape[1])))
    x = F.pad(x, (0, 0, 2**L_log2-x.shape[1], 0, 0, 0))
    N, L, D = x.shape  # left padded sequence shape
    # 2. recursive split
    h = torch.matmul(x, B)
    for i in range(1, L_log2+1):
        l = 2 ** i  # length of subsequences
        h = h.reshape(N*L//l, l, D)
        # 3. parallel forward pass
        h1, h2 = h[:, :l//2], h[:, l//2:]
        if i > 1:  # compute [La, La^2, ..., La^l]
            La = torch.cat((La, La * La[-1]), 0)
        h2 = h2 + La * h1[:, -1:]  # linear recurrence
        h = torch.cat([h1, h2], 1)
    return h
\end{lstlisting}

\begin{figure*}[t]
    \centering
    \includegraphics[trim=2cm 4.2cm 2cm 4cm, clip, width=0.83\linewidth]{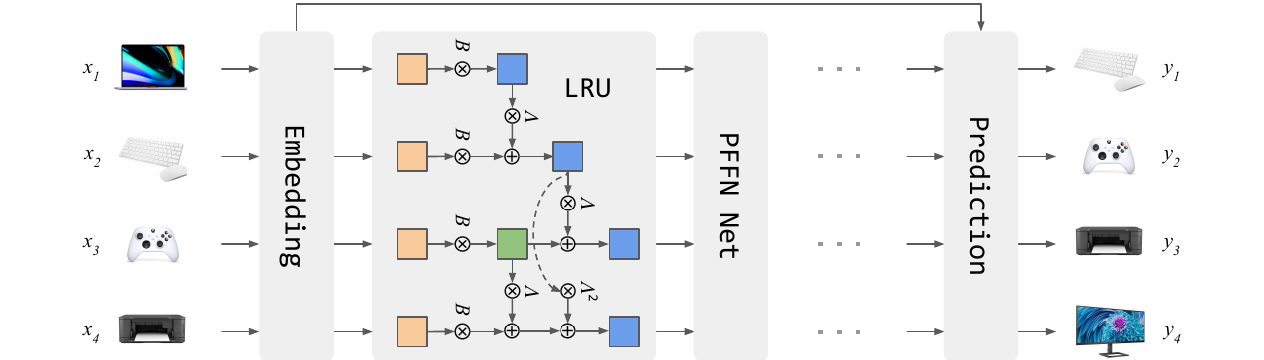}
    \caption{The overall architecture of the proposed \ours.}
    \label{fig:method}
\end{figure*}

\xhdr{1. Sequence Padding}: Given input sequence $x$ of length t, we first pad the sequence length to the power of two (i.e., $2^2, 2^3, \ldots$). The reason for such padding is to enable recursive split of the input sequences until reaching the shortest length of two (i.e., $[x_1, x_2], [x_3, x_4], [x_5, x_6], \ldots$), which maximizes the performance of our recursive parallelization algorithm.

\xhdr{2. Recursive Split}: With the padded sequence of length $2^k$, we multiply $x$ with $B$ and perform $k$ forward passes. In the $i$-th step, the input sequence is split into $2^{k-i}$ subsequences, each of length $2^i$ (i.e., $[x_1, x_2, \ldots, x_{2^i}], [x_{2^i+1}, x_{2^i+2}, \ldots, x_{2*2^i}], \ldots$). The subsequences are processed in parallel to perform linear recurrence (see next step), followed by restoring the original sequence. Consequently, the time complexity reduces with no additional space required.

\xhdr{3. Parallel Forward Pass}: For each input $x = [x_1, x_2, \ldots, x_{2l}]$ of length $2l$ ($l \geq 1$), we further split the input into two equal subsequences of length $l$, namely $[x_1, x_2, \ldots, x_l]$ and $[x_{l+1}, x_{l+2}, \ldots, x_{2l}]$. The last element of the first subsequence (i.e., $x_l$) is multiplied with $[\Lambda, \Lambda^2, \ldots, \Lambda^l]$ respectively, and element-wise added to the second subsequence. In other words, the second subsequence becomes $[x_{l+1} + \Lambda x_l, x_{l+2} + \Lambda^2 x_l, \ldots, x_{2l} + \Lambda^l x_l]$. Then, we restore the original sequence shape: $[x_1, x_2, \ldots, x_l, x_{l+1} + \Lambda x_l, x_{l+2} + \Lambda^2 x_l, \ldots, x_{2l} + \Lambda^l x_l]$. Such forward passes are performed in parallel for all subsequences in each of the $k$ forward passes in the recursive split step.

Recursive parallelization can be perform regardless of input length and hidden dimension, and is designed for full-length training and inference. We provide an implementation of recursive parallelization with PyTorch-like pseudocode in \Cref{lst:parallel}, with $\mathtt{x}$, $\mathtt{B}$ and $\mathtt{La}$ representing input $x$, parameters $B$ and $\Lambda$. The above steps follow the divide-and-conquer principle to break down long sequences into subsequences recursively, such that linear recurrence can be computed on the subsequences in parallel. By applying recursive parallelization, we reduce the number of forward passes to $\log_2(L)$ for input sequence of length $L$ (e.g., three passes for an eight-element sequence, see \Cref{fig:parallel}), which significantly improves the time efficiency for both training and inference in \ours.

\subsection{Overall \ours Model}

In this section, we introduce the overall architecture of the proposed \ours. The proposed model comprises of: (1)~embedding module; (2)~LRU block with position-wise feed-forward network (PFFN) and (3)~prediction layer. The overall model is illustrated in \Cref{fig:method}, we describe the details of each module in the following.

\xhdr{Embedding Module.} Similar to existing methods, we use a learnable matrix $E \in \mathbb{R}^{|\mathcal{I}| \times H_{\mathrm{in}}}$ to transform the discrete item IDs in $\mathcal{I}$ from the input sequence to the high-dimensional embedding space $\mathbb{R}^{H_{\mathrm{in}}}$. For sequences of different lengths, we perform left padding to the power of two for parallelization, as demonstrated in \Cref{sec:parallelization}. Layer normalization ($\mathtt{LayerNorm}$) is performed after the embedding retrieval. Hence, for input sequence $x = [x_1, x_2, ..., x_t]$, we denote the embedding module with $\mathtt{Embed}$:
\begin{equation}
  \begin{aligned}
    \mathtt{Embed}(E, x) &= \mathtt{LayerNorm}([E_{x_1}, E_{x_2}, \ldots, E_{x_t}]), \\
    \mathtt{LayerNorm}(x) &= \alpha * \frac{x - \mu}{\sqrt{\sigma^2 + \epsilon}} + \beta,
  \label{eq:embed}
  \end{aligned}
\end{equation}
where $\alpha$ and $\beta$ are learnable rescaling factors, $\mu$ and $\sigma$ represents the mean and standard deviation of the input, $\epsilon$ is added to the denominator in $\mathtt{LayerNorm}$ for numerical stability.

\xhdr{LRU Block.} As in \Cref{sec:lru}, we use $\mathtt{LRU}$ to describe the linear recurrence operation in \ours. We additionally apply layer normalization on the output hidden features of $\mathtt{LRU}$ and denote the process with $\mathtt{LRUNorm}$. Nevertheless, $\mathtt{LRU}$ sacrifices non-linearity in the recurrence operation for improved performance and efficiency. To compensate for the absence of non-linearity, we leverage position-wise feed-forward network (PFFN) to improve the modeling of user actions in the hidden dimension. In particular, we use $\mathtt{PFFN}$ to describe the two-layer MLP network:
\begin{equation}
    \mathtt{PFFN}(x) = \mathtt{GELU}(W^{(2)} \mathtt{GELU}(W^{(1)} x + b^{(1)}) + b^{(2)}),
    \label{eq:pffn}
\end{equation}
where $W^{(1)} \in \mathbb{R}^{4H \times H}, W^{(2)} \in \mathbb{R}^{H \times 4H}, b^{(1)} \in \mathbb{R}^{4H}, b^{(2)} \in \mathbb{R}^{H}$ are the parameters for the two-layer MLP, and $\mathtt{GELU}$ refers to the GELU activation. We additionally introduce sublayer connection ($\mathtt{SubLayer}$) with both layer normalization and residual connection on the PFFN net to improve the recommendation performance and training dynamics. In short, we leverage linear recurrence unit to efficiently process sequential input. PFFN, layer normalization and residual connections are additionally introduced to improve the modeling of non-linear transition patterns in \ours:
\begin{equation}
  \begin{aligned}
    \mathtt{LRUNorm}(x) &= \mathtt{LayerNorm}(\mathtt{LRU}(x)), \\
    \mathtt{SubLayer}(\mathtt{PFFN}, x) &= \mathtt{LayerNorm}(\mathtt{PFFN}(x) + x).
    \label{eq:lru-with-pffn}
  \end{aligned}
\end{equation}
In our experiments, we stack two LRU blocks following~\cite{kang2018self, sun2019bert4rec}.

\xhdr{Prediction Layer.} Given hidden features $h_t$ at time step $t$ from the previous LRU block, we compute the scores over $\mathcal{I}$ for next-item prediction via the $\mathtt{Pred}$ function:
\begin{equation}
    \mathtt{Pred}(h_t) = E h_t + b^o,
    \label{eq:prediction}
\end{equation}
where $E$ is the embedding matrix from the embedding module and $b^o \in \mathbb{R}^{| \mathcal{I} |}$ is an additional bias term. Thanks to the non-linearity introduced in the LRU block, dot product between embedding features and $h_t$ can capture non-trivial patterns despite utilizing shared item features $E$. Additionally, the shared $E$ significantly reduces the model size, while effectively alleviates overfitting in \ours.

\section{Discussion}

\xhdr{Why does linear recurrence perform well in sequential recommendation?}
We have two explanations for the improvements of \ours: (1)~Linear recurrence by design assigns higher weights to recent items, thus it is effective for modeling recommendation data that emphasizes recent interactions. To examine this, we inspect the average $|\lambda|$ values (i.e., $e^{-\nu}$) on different sequence lengths. Notice that higher $|\lambda|$ values suggest the inclusion of more history information. With $\sim$200 sequence length (i.e., ML-1M), we observe $\sim$0.4 for both LRU blocks. The values reduce to 0.25 and 0.31 for $\sim$10 sequence length in Beauty (see \Cref{sec:xlong}). As $|\lambda|$ is initialized close to 1, the low values indicate high emphasis on recent items, thus providing a solid justification for the recurrence design in \ours. (2)~PFFN and hierarchical LRU blocks further improve the modeling of non-trivial transition patterns in long-range dependencies and dense datasets. For example, PFFN and stacking LRU block causes up to $11.24\%$ and $3.43\%$ performance improvements on long sequences (ML-1M), while contributing less than $1\%$ improvements on the sparse Beauty data (see \Cref{sec:ablation}). Hence, the combination of additional non-linearity and hierarchical linear recurrence plays a vital role in the overall performance of \ours.

\xhdr{How does the complexity of \ours compare to other models?} We use $H$ and $L$ to represent the model hidden dimension and input sequence length, $\mathcal{I}$ and $\mathcal{U}$ to represent the items and users. The quantity of the learnable parameters in the proposed \ours is comparable to RNN-based sequential recommenders, which have $\mathcal{O}(|\mathcal{I}|H + H^2)$ space complexity. In comparison, factorization models have $\mathcal{O}((|\mathcal{I}| + |\mathcal{U}|)H)$ and transformer models have $\mathcal{O}((L + |\mathcal{I}|)H + H^2)$ space complexity. Given the low hidden dimension size ($H = 64$ in our experiments), the space complexity of \ours is smaller in size and does not grow with increasing user numbers. Moreover, we use identity matrix as $D$ and decompose $A = P \Lambda P^{-1}$ ($P$ is integrated in $B$ and $C$ matrices), which leads to additional parameter reduction in \ours. For time complexity, \ours achieves $\mathcal{O}(\mathrm{log}(L) H^2)$ for full-length input sequence as a result of our recursive parallelization design. In contrast, typical RNN-based recommenders require $\mathcal{O}(L H^2)$ in time and the original transformer-based recommenders require $\mathcal{O}(L^2 H + L H^2)$. Like RNNs, \ours additionally support incremental inference with only $\mathcal{O}(H^2)$ complexity. We demonstrate the advantages of \ours on both performance and efficiency in \Cref{sec:experiment}.

\begin{table}[t]
    \small
    \centering
    \begin{tabular}{lrrcS[table-text-alignment=right,table-format=3.1]c}
    \toprule
    \textbf{Datasets} & \textbf{Users} & \textbf{Items} & \textbf{Interact.} & \textbf{Lenth} & \textbf{Density} \\ \midrule
    \textbf{ML-1M}    & 6,040          & 3,416          & 1M                 & 165.5             & 5e-2             \\
    \textbf{Beauty}   & 52,204         & 57,289         & 398K               & 7.6               & 1e-4             \\
    \textbf{Video}    & 31,013         & 23,715         & 287K               & 9.3               & 4e-4             \\
    \textbf{Sports}   & 83,970         & 83,728         & 596K               & 7.1               & 8e-5             \\
    \textbf{Steam}    & 334,730        & 13,047         & 3.7M               & 11.0              & 8e-4             \\
    \textbf{XLong}    & 69,691         & 2,122,932      & 66.8M              & 958.8             & 5e-4             \\
    \bottomrule
    \end{tabular}
    \caption{Dataset statistics after preprocessing.}
    \label{tab:dataset}
\end{table}

\begin{table*}[t]
\small
\centering
\begin{tabularx}{1.0\linewidth}{ll*{10}{>{\centering\arraybackslash}X}}
\toprule
\textbf{Dataset}                 & \textbf{Metric}        & \textbf{MF} & \textbf{FISM} & \textbf{FPMC} & \textbf{GRU4Rec} & \textbf{NARM} & \textbf{SASRec} & \textbf{BERT4Rec} & $\textbf{FMLP-Rec}$ & \textbf{LRURec}  & \textbf{Improv.} \\ \midrule
\multirow{4}{*}{\textbf{ML-1M}}  & NDCG@10 $\uparrow$     & 0.02835     & 0.03327       & 0.12030       & 0.15943          & 0.15310       & \ul{0.18293}    & 0.16377           & 0.15295            & \textbf{0.19065} & 4.22\%           \\
                                 & NDCG@20 $\uparrow$     & 0.03991     & 0.04655       & 0.14370       & 0.18735          & 0.17774       & \ul{0.21232}    & 0.19118           & 0.17912            & \textbf{0.21767} & 2.52\%           \\
                                 & Recall@10 $\;\uparrow$ & 0.06290     & 0.07368       & 0.23030       & 0.28317          & 0.27290       & \ul{0.31441}    & 0.30894           & 0.29073            & \textbf{0.32402} & 3.06\%           \\
                                 & Recall@20 $\;\uparrow$ & 0.10910     & 0.12520       & 0.31510       & 0.39378          & 0.37066       & \ul{0.43084}    & 0.41738           & 0.39454            & \textbf{0.43118} & 0.08\%           \\ \cmidrule(l){2-12} 
\multirow{4}{*}{\textbf{Beauty}} & NDCG@10 $\uparrow$     & 0.01193     & 0.01281       & 0.02091       & 0.01997          & 0.01887       & \ul{0.02917}    & 0.02422           & 0.02515            & \textbf{0.03088} & 5.86\%           \\
                                 & NDCG@20 $\uparrow$     & 0.01563     & 0.01621       & 0.02371       & 0.02343          & 0.02239       & \ul{0.03403}    & 0.02937           & 0.02945            & \textbf{0.03560} & 4.61\%           \\
                                 & Recall@10 $\;\uparrow$ & 0.02596     & 0.02614       & 0.03586       & 0.03379          & 0.03312       & \ul{0.05043}    & 0.04679           & 0.04623            & \textbf{0.05284} & 4.78\%           \\
                                 & Recall@20 $\;\uparrow$ & 0.03987     & 0.04008       & 0.04699       & 0.04751          & 0.04708       & \ul{0.06965}    & 0.06727           & 0.06336            & \textbf{0.07161} & 2.81\%           \\ \cmidrule(l){2-12} 
\multirow{4}{*}{\textbf{Video}}  & NDCG@10 $\uparrow$     & 0.01337     & 0.02813       & 0.04318       & 0.05081          & 0.05329       & \ul{0.05829}    & 0.05600           & 0.05294            & \textbf{0.06198} & 6.33\%           \\
                                 & NDCG@20 $\uparrow$     & 0.01722     & 0.03692       & 0.05267       & 0.06200          & 0.06512       & \ul{0.07111}    & 0.06958           & 0.06576            & \textbf{0.07547} & 6.13\%           \\
                                 & Recall@10 $\;\uparrow$ & 0.02634     & 0.05815       & 0.08110       & 0.09512          & 0.09777       & 0.10910         & \ul{0.11252}      & 0.10885            & \textbf{0.11337} & 0.76\%           \\
                                 & Recall@20 $\;\uparrow$ & 0.04052     & 0.09281       & 0.11830       & 0.13959          & 0.14488       & 0.15989         & \ul{0.16640}      & 0.15991            & \textbf{0.16705} & 0.39\%           \\ \cmidrule(l){2-12} 
\multirow{4}{*}{\textbf{Sports}} & NDCG@10 $\uparrow$     & 0.00266     & 0.00810       & 0.01070       & 0.01002          & 0.01117       & 0.01236         & 0.01457           & \ul{0.01546}       & \textbf{0.01815} & 17.43\%          \\
                                 & NDCG@20 $\uparrow$     & 0.00362     & 0.01036       & 0.01257       & 0.01243          & 0.01414       & 0.01520         & 0.01813           & \ul{0.01871}       & \textbf{0.02149} & 14.86\%          \\
                                 & Recall@10 $\;\uparrow$ & 0.00413     & 0.01614       & 0.01936       & 0.01882          & 0.02087       & 0.02294         & 0.02879           & \ul{0.02987}       & \textbf{0.03148} & 5.39\%           \\
                                 & Recall@20 $\;\uparrow$ & 0.00861     & 0.02644       & 0.02643       & 0.02841          & 0.03265       & 0.03428         & \ul{0.04295}      & 0.04276            & \textbf{0.04477} & 4.24\%           \\ \cmidrule(l){2-12} 
\multirow{4}{*}{\textbf{Steam}}  & NDCG@10 $\uparrow$     & 0.06480     & 0.04209       & 0.04377       & 0.06512          & 0.06480       & \ul{0.06759}    & 0.06433           & 0.06081            & \textbf{0.06934} & 2.59\%           \\
                                 & NDCG@20 $\uparrow$     & 0.07996     & 0.05439       & 0.05582       & 0.08050          & 0.07996       & \ul{0.08298}    & 0.08117           & 0.07521            & \textbf{0.08508} & 2.53\%           \\
                                 & Recall@10 $\;\uparrow$ & 0.12166     & 0.08561       & 0.08773       & 0.12204          & 0.12166       & \ul{0.12597}    & 0.12188           & 0.11987            & \textbf{0.12831} & 1.86\%           \\
                                 & Recall@20 $\;\uparrow$ & 0.18199     & 0.13450       & 0.13570       & 0.18319          & 0.18199       & \ul{0.18716}    & 0.18245           & 0.17719            & \textbf{0.19082} & 1.96\%           \\ \bottomrule
\end{tabularx}
\caption{Main performance results, best results are marked in bold, second best results underlined.}
\label{tab:main-results}
\end{table*}

\section{Experiments}
\label{sec:experiment}

\subsection{Experimental Setup}

\xhdr{Datasets.}
\label{datasets}
Our model is evaluated on the following datasets:
\begin{itemize}
    \item \textbf{MovieLens}: A benchmark dataset for movie recommendation, we select the widely used ML-1M here~\cite{harper2015movielens}.
    \item \textbf{Amazon}: A series of datasets with product reviews from Amazon,
    here we select Beauty, Video and Sports~\cite{mcauley2015image, he2016ups}.
    \item \textbf{Steam}: A video game review dataset crawled from Steam, a large online video game distribution platform~\cite{kang2018self}.
    \item \textbf{XLong}: XLong is an Alibaba dataset known for long interaction histories to evaluate lifelong sequential models~\cite{ren2019lifelong}.
\end{itemize}
For preprocessing, we follow~\cite{kang2018self, he2021locker, yue2021black, yue2022defending} to construct input sequences and exclude users and items with fewer than 5 interactions. For maximum sequence length, we adopt 200 for ML-1M, 1000 for XLong and 50 for the rest datasets. We follow the leave-last-out strategy on dataset splitting and use the most recent item as the test set, the second most recent item as the validation set, and the rest items in the sequences as the training set. During testing, we include both training and validation items as input.  We report the dataset statistics after preprocessing in \Cref{tab:dataset}.

\xhdr{Baseline Methods.}
We compare our \ours against multiple baseline methods, which include classic factorization and Markov chain-based methods (e.g. MF, FISM, FPMC), RNN-based models (e.g. GRU4Rec, NARM) as well as transformer- and MLP-based state-of-the-art models (e.g., SASRec, BERT4Rec and FMLP-Rec):
\begin{itemize}
    \item \textbf{MF}: A vanilla factorization model that learns user and item latent representations for next-item prediction~\cite{koren2009matrix}.
    \item \textbf{FISM}: FISM does not explicitly model user preference and predicts next interaction via item-to-item similarity~\cite{kabbur2013fism}.
    \item \textbf{FPMC}: FPMC is a matrix factorization model and uses Markov chains to capture user transition patterns~\cite{rendle2010factorizing}.
    \item \textbf{GRU4Rec}: A classic sequential recommender that models user-item interactions using a GRU-based model~\cite{hidasi2015session, hidasi2018recurrent}.
    \item \textbf{NARM}: Also a RNN-based sequential recommender with local and global user modeling for next-item prediction~\cite{li2017neural}.
    \item \textbf{SASRec}: The first unidirectional transformer-based sequential recommender, SASRec leverages unidirectional self-attention to capture user-item transition patterns~\cite{kang2018self}. 
    \item \textbf{BERT4Rec}: A bidirectional transformer encoder architecture for sequential recommendation. BERT4Rec is learnt via predicting a random proportion of masked items~\cite{sun2019bert4rec}.
    \item \textbf{FMLP-Rec}: A filter-based all-MLP model that learns in the complex domain using fast Fourier transformation~\cite{zhou2022filter}. 
\end{itemize}

\xhdr{Evaluation.}
For evaluation results, we select models with the best validation Recall@10 scores in training to perform prediction on the test sets. The models are evaluated using Recall@k and NDCG@k metrics, and with $k \in \{10,20\}$. The predicted items are ranked against all items in the dataset to compute the final scores.

\xhdr{Implementation Details.}
\label{Implementation Details}
For the baseline methods, we refer to the original papers for implementation. We train all models using cross entropy loss with AdamW optimizer using the static learning rate of 1e-3. During training, we use a batch size of 128 (32 for XLong) and set the maximum epoch to be 500, validation is performed every 500 to 2000 iterations depending on the data size. Early stopping is triggered if validation Recall@10 does not improve in 10 consecutive validation rounds. We perform grid search for hyperparameters, with weight decay from [$0$, 1e-6, 1e-4, 1e-2] and dropout rate from [$0.2, 0.4, 0.6, 0.8$]. Hyperparameters not mentioned above were used as reported in the original implementation\footnote{Notice that our preprocessing differs from the k-core method and leads to more realistic and sparser distributions. As such, the reported numbers should not be directly compared with those found in works that employ k-core preprocessing.}.

\begin{table}[t]
\small
\centering
\begin{tabular}{@{}lcccc@{}}
\toprule
\multirow{2}{*}{\textbf{Methods}} & \multicolumn{4}{c}{\textbf{XLong}}                                                     \\ \cmidrule(l){2-5} 
                                  & NDCG@10 $\uparrow$ & Recall@10 $\uparrow$ & NDCG@20 $\uparrow$  & Recall@20 $\uparrow$ \\ \midrule
\textbf{SASRec}                   & \ul{0.30949} & \ul{0.49348}               & \ul{0.33531} & \ul{0.59527}                \\
\textbf{BERT4Rec}                 & 0.23426 & 0.41536                         & 0.27293 & 0.56860                          \\
\textbf{\ours}                    & \textbf{0.34867} & \textbf{0.52688}       & \textbf{0.37174} & \textbf{0.61800}        \\ \bottomrule
\end{tabular}
\caption{Long-range modeling performance results, best results are marked in bold, second best results underlined.}
\label{tab:xlong}
\end{table}

\subsection{Overall Performance}
Our main performance results are reported in \Cref{tab:main-results}. Here, rows represent the dataset and metric, and the columns represent each of the methods, we mark the best results in bold and underline the second best results. We also compute the relative improvement of \ours compared to the best-performing baseline method (i.e., Improv.). We observe: (1)~\ours consistently outperforms baseline methods across all metrics and datasets, with an average performance improvement of $4.39\%$ compared to the second best method. Despite the efficient and light-weight design, the performance gains of \ours can go up to over $10\%$ depending on the data distribution (e.g., $17.43\%$ on NDCG@10 in Sports). (2)~The performance gains of \ours are more pronounced on sparse datasets, while being comparatively modest on dense datasets. For example, \ours achieves $1.32\%$ average improvements on the relatively dense ML-1M. The improvement is much more significant on the sparse Sports dataset with average gains of $10.48\%$, suggesting the substantial benefits of \ours on sparse data. (3)~In contrast to recall, \ours demonstrates better ranking performance. For instance, there is a noteworthy increase of $6.73\%$ in the average NDCG@10 scores with \ours, while the relative improvement on Recall@10 is slightly lower at $3.08\%$. Overall, we find \ours performs particularly well on sparse data and shows significantly improved ranking performance compared to the baseline methods. \ours also achieves consistent performance improvements in all scenarios, suggesting the effectiveness of \ours regardless of data domains.

\begin{table}[t]
\small
\centering
\begin{tabular}{@{}lcccccc@{}}
\toprule
\textbf{}                    & \textbf{ML-1M} & \textbf{Beauty} & \textbf{Video} & \textbf{Sports} & \textbf{Steam} & \textbf{XLong} \\ \midrule
\textbf{$|\lambda|$ Block 1} & 0.3927         & 0.2540          & 0.2501         & 0.3675          & 0.3382         & 0.7720         \\
\textbf{$|\lambda|$ Block 2} & 0.4216         & 0.3099          & 0.3047         & 0.4287          & 0.3687         & 0.8045         \\ \bottomrule
\end{tabular}
\caption{Average $|\lambda|$ values of each LRU block, higher $|\lambda|$ indicates the incorporation of increased history information.}
\label{tab:lambda}
\end{table}

\begin{table*}[t]
\small
\centering
\begin{tabular}{@{}lcccccc@{}}
\toprule
\multirow{2}{*}{\textbf{Variants}}                         & \multirow{2}{*}{\textbf{Metric}} & \textbf{ML-1M}                      & \textbf{Beauty}                     & \textbf{Video}                      & \textbf{Sports}                     & \textbf{Steam}                      \\ \cmidrule(l){3-7} 
                                                           &                                  & NDCG $\uparrow$ / Recall $\uparrow$ & NDCG $\uparrow$ / Recall $\uparrow$ & NDCG $\uparrow$ / Recall $\uparrow$ & NDCG $\uparrow$ / Recall $\uparrow$ & NDCG $\uparrow$ / Recall $\uparrow$ \\ \midrule
\multirow{2}{*}{\textbf{\ours}}                            & @10                              & \textbf{0.19298} / \ul{0.32636}     & \ul{0.03088} / \textbf{0.05284}     & \textbf{0.06198} / \textbf{0.11337} & \textbf{0.01815} / \textbf{0.03148} & \ul{0.06934} / \ul{0.12831}         \\
                                                           & @20                              & \textbf{0.21990} / \textbf{0.43313} & \ul{0.03560} / \textbf{0.07161}     & \textbf{0.07547} / \textbf{0.16705} & \textbf{0.02149} / \textbf{0.04477} & \ul{0.08508} / \ul{0.19082}         \\ \midrule
\multirow{2}{*}{(1) \textbf{\ours} w/o \texttt{LayerNorm}} & @10                              & \ul{0.19242} / \textbf{0.32661}     & 0.01562 / 0.02904                   & 0.05222 / 0.09435                   & 0.01072 / 0.02008                   & 0.06793 / 0.12691                   \\
                                                           & @20                              & \ul{0.21751} / \ul{0.42584}         & 0.01929 / 0.04364                   & 0.06430 / 0.14232                   & 0.01348 / 0.03106                   & 0.08360 / 0.18920                   \\ \cmidrule(l){2-7} 
\multirow{2}{*}{(2) \textbf{\ours} w/o \texttt{Residual}}  & @10                              & 0.18038 / 0.31185                   & 0.00932 / 0.01924                   & 0.02827 / 0.05503                   & 0.00580 / 0.01205                   & 0.06138 / 0.11639                   \\
                                                           & @20                              & 0.20606 / 0.41369                   & 0.01244 / 0.03171                   & 0.03552 / 0.08394                   & 0.00763 / 0.01934                   & 0.07648 / 0.17647                   \\ \cmidrule(l){2-7} 
\multirow{2}{*}{(3) \textbf{\ours} w/o \texttt{PFFN}}      & @10                              & 0.17319 / 0.29318                   & 0.03075 / 0.05249                   & 0.06019 / 0.11177                   & \ul{0.01769} / 0.03026              & 0.06565 / 0.12251                   \\
                                                           & @20                              & 0.19768 / 0.39024                   & 0.03556 / \ul{0.07157}              & 0.07356 / 0.16496                   & \ul{0.02109} / \ul{0.04380}         & 0.08098 / 0.18348                   \\ \midrule
\multirow{2}{*}{(4) \textbf{Backbone}: LRU (1 Layer)}      & @10                              & 0.18777 / 0.31489                   & \textbf{0.03120} / \ul{0.05276}     & \ul{0.06017} / \ul{0.11102}         & 0.01742 / 0.03012                   & 0.06784 / 0.12563                   \\
                                                           & @20                              & 0.21330 / 0.41569                   & \textbf{0.03580} / 0.07104          & \ul{0.07350} / \ul{0.16400}         & 0.02062 / 0.04279                   & 0.08323 / 0.18684                   \\ \cmidrule(l){2-7}
\multirow{2}{*}{(6) \textbf{Backbone}: RWKV}               & @10                              & 0.15942 / 0.28411                   & 0.02620 / 0.04653                   & 0.04913 / 0.09400                   & 0.01663 / \ul{0.03071}              & 0.04744 / 0.09081                   \\
                                                           & @20                              & 0.18549 / 0.39112                   & 0.03089 / 0.06511                   & 0.05983 / 0.13643                   & 0.01962 / 0.04255                   & 0.05915 / 0.13737                   \\ \cmidrule(l){2-7} 
\multirow{2}{*}{(5) \textbf{Backbone}: S4}                 & @10                              & 0.16460 / 0.28158                   & 0.02260 / 0.03671                   & 0.05039 / 0.09203                   & 0.01120 / 0.01907                   & \textbf{0.06945} / \textbf{0.13044} \\
                                                           & @20                              & 0.18972 / 0.38122                   & 0.02584 / 0.04956                   & 0.06212 / 0.13869                   & 0.01315 / 0.02680                   & \textbf{0.08573} / \textbf{0.19515} \\ \bottomrule

\end{tabular}
\caption{Ablation results, best results are marked in bold, second best results underlined.}
\label{tab:ablation-results}
\end{table*}

\subsection{Long-Range Modeling Performance}
\label{sec:xlong}
To examine the performance of \ours on long-range dependencies, we additionally experiment on XLong: a large-scale dataset with $\sim$1k sequence length. Due to scalability issues, we only experiment on selected state-of-the-art baselines (SASRec and BERT4Rec) along with \ours\footnote{The demand for intricate sequence processing and augmentation renders the training of XLong computationally infeasible for RNN-based models and FMLP-Rec.}. We also reduce the hyperparameter search to [$0$, 1e-2] for weight decay and [$0.2, 0.4$] for dropout rate. We improve the training efficiency on XLong by randomly sampling 100 negative items to compute loss and update model in training. For evaluation, we randomly sample 10k negative items compute the metrics, the experiment results are reported in \Cref{tab:xlong}. Analogous to the main results, \ours outperforms baseline methods by a considerable margin, indicating the effectiveness of \ours even for long-range dependencies. For example, \ours achieves $8.53\%$ average improvements across all metrics compared to the best-performing baseline SASRec. In addition, we evaluate the weights of history information in \ours by computing the average $|\lambda|$ in each of the LRU blocks. The $|\lambda|$ values for all datasets are reported in \Cref{tab:lambda}. As expected, we observe high $|\lambda|$ values for long sequences (e.g., $\sim$0.8 for XLong), whereas for short sequences, the $|\lambda|$ values are significantly lower (e.g., $\sim$0.3 for Beauty and Video). Interestingly, we observe relatively high $|\lambda|$ on Sports despite its short sequence length, which may explain for the significant performance improvement of \ours (up to $17.43\%$) in the main results.

\subsection{Ablation Studies}
\label{sec:ablation}
We perform a series of ablations to demonstrate the effectiveness of the proposed components in \ours. In particular, we remove the layer normalization, residual connection and PFFN net respectively and evaluate the performance changes. We additionally reduce the LRU blocks in \ours and adopt two efficient sequence modeling methods, S4 and RWKV, to replace LRU~\cite{gu2021efficiently, peng2023rwkv}. The ablation results are reported in \Cref{tab:ablation-results}, with rows representing the ablation variant and columns representing the datasets. We observe the following on the ablation of \ours components: (1)~All components contribute to the overall performance of \ours. For example, removing PFFN results in an average performance drop of $11.24\%$ on ML-1M. (2)~The components contribute differently depending on the datasets. For instance, layer normalization and residual connection contribute significantly to the performance on sparse datasets (e.g., with $82.07\%$ and over $100\%$ average gains on Beauty). In contrast, PFFN improves the modeling of non-linear transition patterns, and thereby further enhances the performance on dense datasets like ML-1M. By additionally switching the backbone of our \ours, we notice: (1)~The overall best performing variant is still our reduced one-layer \ours, demonstrating the effectiveness of the proposed architecture. On average, the one-layer \ours outperforms the second-best backbone variant by $7.70\%$ on Recall@10. (2)~Surprisingly, the S4 variant performs the best on the Steam dataset, which may be attributed to the combination of the linear design of S4 and the imbalanced item popularity in Steam. Overall, the ablation results suggest that all proposed components and the carefully designed architecture in \ours are effective for sequential recommendation across various data scenarios.

\subsection{Model Efficiency}
To further demonstrate the advantages of our design, we study the efficiency of \ours against two representative baselines: RNN-based GRU4Rec and transformer-based SASRec. We illustrate validation Recall@10 curves during training in \Cref{fig:training-inference} (left), with the horizontal axis representing training steps and the vertical axis representing Recall@10 scores. Owning to the linear recurrence design and improved training dynamics, \ours converges significantly faster and triggers early stopping at $\sim$23k training steps, compared to over 50k of SASRec and over 80k of GRU4Rec. Aside from training, \ours also demonstrates substantial advantages with incremental inference that drastically reduces latency for real-time recommendation, as illustrated in \Cref{fig:training-inference} (right). Here, we perform batched inference and keep extending input length after each prediction step, we visualize the cumulative prediction time for a total of 2048 steps. Given hidden states and current input elements, we observe almost linear correlation between the cumulative computing time and increasing input steps on both GRU4Rec and \ours. Moreover, the throughput capacity of \ours is independent of the the sequence length in incremental inference. For example, \ours can achieve $\times$7.3 increased input examples compared to SASRec with maximum sequence length of 50. In summary, the results suggest that \ours has the benefits in training parallelization and performance like transformer models, while retaining the advantage of incremental inference from RNNs.

\begin{figure}[t]
    \centering
    \includegraphics[trim=1.1cm 0 2.6cm 0.9cm, clip, width=1.0\linewidth]{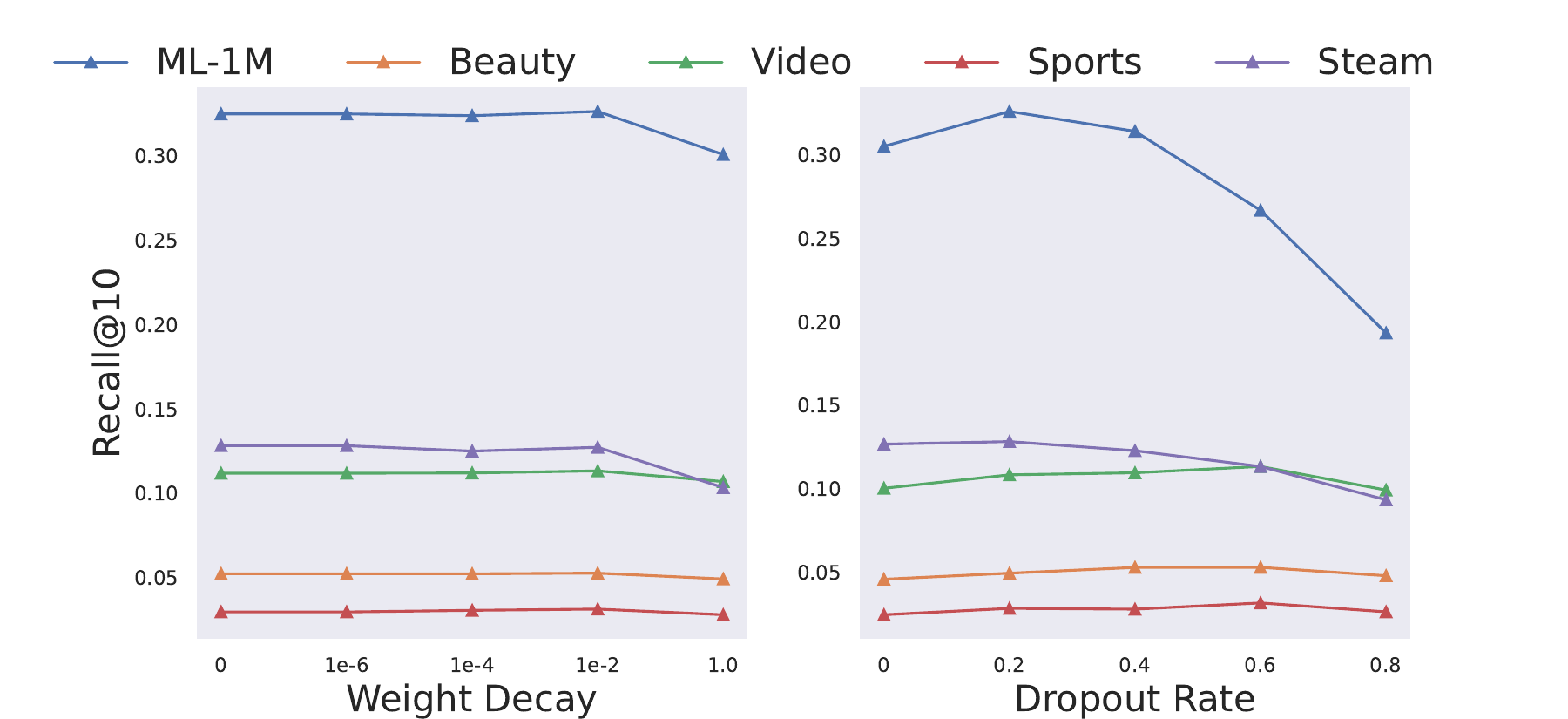}
    \caption{Hyperparameter sensitivity of \ours.}
    \label{fig:sensitivity}
\end{figure}

\subsection{Hyperparameter Sensitivity}
\label{sec:sensitivity}

We evaluate the hyperparameter sensitivity of \ours. In particular, we vary weight decay and dropout values to evaluate the trained models with the best validation performance. \Cref{fig:sensitivity} compares the performance with different hyperparameters, with the x-axis stands for the varying values and y-axis stands for Recall@10 scores. For weight decay, we observe minor changes with increasing penalty strength, the performance remains robust until increasing weight decay to 1. For varying dropout rates, we observe different performance changes depending on the datasets. For dense datasets (e.g., ML-1M), the best performance is achieved at 0.2 and then consistently reduces with increasing dropout rates. Unlike dense datasets, Recall@10 performance peaks at 0.6 dropout rate on sparse datasets like Video and Beauty. Overall, the performance of \ours is quite robust with varying weight decay values, while dropout rates should be carefully selected for optimal performance.

\section{Conclusion}
In this paper, we propose a novel sequential recommender called \ours. The proposed model introduces: (1)~linear recurrence with matrix diagonalization to efficiently capture user transition patterns; and (2)~a recursive parallelization framework that significantly accelerates  training and inference. Moreover, \ours is designed with a series of improvements for optimal recommendation performance and inference efficiency. We demonstrate the effectiveness and efficiency of \ours by performing extensive experiments on multiple real-world datasets, where \ours consistently achieves superior results over state-of-the-art baselines.

\newpage
\bibliographystyle{ACM-Reference-Format}
\bibliography{reference}

\end{document}